\documentstyle[12pt,aaspp4]{article}

\begin{document}

\title{OBSERVATION OF MULTI-TeV GAMMA RAYS FROM THE CRAB NEBULA USING THE TIBET AIR SHOWER ARRAY}

\author{ M.~Amenomori\altaffilmark{1}, S.~Ayabe\altaffilmark{2}, 
        P.Y.~Cao\altaffilmark{3},       Danzengluobu\altaffilmark{4},
        L.K.~Ding\altaffilmark{5},     Z.Y.~Feng\altaffilmark{6}, 
        Y.~Fu\altaffilmark{3},          H.W.~Guo\altaffilmark{4}, 
        M.~He\altaffilmark{3},          K.~Hibino\altaffilmark{7},
        N.~Hotta\altaffilmark{8},       Q.~Huang\altaffilmark{6},
        A.X.~Huo\altaffilmark{5},       K.~Izu\altaffilmark{9},
        H.Y.~Jia\altaffilmark{6},       F.~Kajino\altaffilmark{10}, 
        K.~Kasahara\altaffilmark{11},   Y.~Katayose\altaffilmark{9},
        Labaciren\altaffilmark{4},      J.Y.~Li\altaffilmark{3},
        H.~Lu\altaffilmark{5},          S.L.~Lu\altaffilmark{5},
        G.X.~Luo\altaffilmark{5},       X.R.~Meng\altaffilmark{4}, 
        K.~Mizutani\altaffilmark{2},     J.~Mu\altaffilmark{12}, 
        H.~Nanjo\altaffilmark{1},       M.~Nishizawa\altaffilmark{13},
        M.~Ohnishi\altaffilmark{9},     I.~Ohta\altaffilmark{8},
        T.~Ouchi\altaffilmark{7},       J.R.~Ren\altaffilmark{5},
        T.~Saito\altaffilmark{14},      M.~Sakata\altaffilmark{10},
        T.~Sasaki\altaffilmark{10},     Z.Z.~Shi\altaffilmark{5},
        M.~Shibata\altaffilmark{15},    A.~Shiomi\altaffilmark{9}, 
        T.~Shirai\altaffilmark{7},      H.~Sugimoto\altaffilmark{16},
        K.~Taira\altaffilmark{16},      Y.H.~Tan\altaffilmark{5},
        N.~Tateyama\altaffilmark{7},    S.~Torii\altaffilmark{7}, 
        T.~Utsugi\altaffilmark{2},      C.R.~Wang\altaffilmark{3},
        H.~Wang\altaffilmark{5},        
        X.W.~Xu\altaffilmark{5},        Y.~Yamamoto\altaffilmark{10},
        G.C.~Yu\altaffilmark{6},        A.F.~Yuan\altaffilmark{4}, 
        T.~Yuda\altaffilmark{9},        C.S.~Zhang\altaffilmark{5},
        H.M.~Zhang\altaffilmark{5},     J.L.~Zhang\altaffilmark{5},
        N.J.~Zhang\altaffilmark{3},     X.Y.~Zhang\altaffilmark{3}, 
        Zhaxisangzhu\altaffilmark{4},   Zhaxiciren\altaffilmark{4},
        and  W.D.~Zhou\altaffilmark{12}
         (The Tibet AS${\bf \gamma}$ Collaboration)}

\altaffiltext{1}{ Department of Physics, Hirosaki University, Hirosaki 036-8561, Japan}
\altaffiltext{2}{ Department of Physics, Saitama University, Urawa 338-8570, Japan}
\altaffiltext{3}{ Department of Physics, Shangdong University, Jinan 250100, China}
\altaffiltext{4}{  Department of Mathematics and Physics, Tibet University, Lhasa 850000, China}
\altaffiltext{5}{ Institute of High Energy Physics, Academia Sinica, Beijing 100039, China}
\altaffiltext{6}{  Department of Physics, South West Jiaotong University, Chengdu 610031, China}
\altaffiltext{7}{  Faculty of Engineering, Kanagawa University, Yokohama 221-8686, Japan}
\altaffiltext{8}{  Faculty of Education, Utsunomiya University, Utsunomiya 321-8505, Japan}
\altaffiltext{9}{  Institute for Cosmic Ray Research, University of Tokyo, Tanashi 188-8502, Japan}
\altaffiltext{10}{ Department of Physics, Konan University, Kobe 658-8501, Japan}
\altaffiltext{11}{ Faculty of Systems Engineering, Shibaura Institute of Technology, Omiya 330-8570, Japan}
\altaffiltext{12}{ Department of Physics, Yunnan University, Kunming 650091, China}
\altaffiltext{13}{ National Center for Science Information Systems, Tokyo 112-8640, Japan}
\altaffiltext{14}{ Tokyo Metropolitan College of Aeronautical Engineering, Tokyo 116-0003, Japan}
\altaffiltext{15}{ Faculty of Engineering, Yokohama National University, Yokohama 240-0067, Japan}
\altaffiltext{16}{ Shonan Institute of Technology, Fujisawa 251-8511, Japan}

\begin{abstract}
The Tibet experiment, operating at Yangbajing (4,300 m above sea level), is the lowest energy 
air shower array and the new high density array constructed in 1996  has sensitivity 
to $\gamma$-ray air showers at energies as low as 3 TeV.
With this new array, the Crab Nebula was observed in multi-TeV $\gamma$-rays
and a signal was detected at the 5.5 $\sigma$ level. We also obtained the energy spectrum of
 $\gamma$-rays in the energy region above 3 TeV  which partially overlaps those observed with 
imaging atmospheric  Cherenkov telescopes.  The Crab spectrum observed in this
energy region  can be represented  by the power-law fit,
$ dJ(E)/dE = (4.61 \pm 0.90) \times 10^{-12}(E/3~TeV)^{-2.62 \pm 0.17} 
{\hspace{2mm}\rm cm^{-2} s^{-1}{TeV^{-1}}} $. 
This is the first observation of $\gamma$-ray signals from  point sources with a 
conventional air shower array using scintillation detectors.

\end{abstract}


\section{INTRODUCTION}

  The Crab Nebula has been well studied with atmospheric Cherenkov telescopes (ACT)
and has been established as a standard source in  TeV astronomy. The Crab was the first 
predicted source of TeV emission (\cite{gou65}) and also the first TeV source to be 
detected at high significance with the advent of imaging ACTs (\cite{week89}).
The observed spectra so far reported (\cite{ong98}) seem to extend  up to at least 
several 10 TeV  and  to be nearly consistent with a Synchrotron-Self Compton (SSC)
 model based on the pulsar wind model of plerions (\cite{dej92}). In this model, 
it is likely that  electrons are accelerated  up to ultra-high energies (UHE) 
extending to somewhat above $10^{15}$ eV  in the pulsar wind shock (\cite{gai89}),
 emitting soft photons by a synchrotron process.
The emission of TeV energies from the Crab can then be  explained as the
inverse-Compton (IC) scattering of such relativistic  electrons with soft photons,
including infrared radiation from dust and  the 2.7K microwave background. The 
IC $\gamma$-ray spectrum depends on the parameter, $\sigma$, which is defined as 
the ratio of the magnetic field to the  particle energy density in the wind, and 
thus on the magnetic field distribution in the nebula.

Measurements of the Crab flux done so far  in the  TeV range  suggest that
the wind is strongly particle dominated and $\sigma$ takes a value larger than
 $\sim$ 0.003, but the flux becomes relatively insensitive to higher values of 
$\sigma$ (\cite{dej96}). A more accurate determination of $\sigma$ is  provided 
by measurements above 10 TeV, which begin to probe  the inner parts of the nebula 
near the shock. The IC $\gamma$-ray spectrum, however,  will  steepen at high energies, 
say higher than 10 TeV since UHE electrons rapidly dissipate their energy 
by the synchrotron radiation. Therefore, the contribution of  $\pi^0$-decay $\gamma$-rays  
becomes more important at high energies (\cite{atoy96}). If the protons
 are accelerated up to very high energies with a hard spectrum in the nebula,
 the $\pi^0$-decay  contribution  would become significant at multi-TeV, or even dominating in the 
energy region higher than several 10 TeV.
Therefore, an accurate measurement of the  $\gamma$-ray spectrum from the Crab  in the high 
energy region can set major constraints on the theoretical explanations and may provide
vital information about the proton acceleration in SNRs, which are the most promising candidate
sources of cosmic rays in our galaxy.  

Until now, using ACTs, several  groups have succeeded in  detecting the Crab with
very high significance  in the TeV energy region, but the systematic uncertainties 
in the flux estimates  remain large. The spectrum, measured by several 
groups (\cite{bhat97}),  differs in power-law slope as well as in absolute flux 
; The derivation of an accurate $\gamma$-ray flux from this source may prove more 
difficult, despite continual improvements in sensitivity of the detection technique. 
Measurements using a different technique will be of great importance to understand the 
possible biases and errors involved in the Cherenkov observations. 

 The Tibet experiment (\cite{ame92}), at an altitude of 4,300 m is the lowest 
energy air shower array and has sensitivity to $\gamma$-ray air showers at energies as 
low as 3 TeV (\cite{ame97}). Thus, observations by the Tibet experiment, using well 
established air shower technique,   partially overlap those of Cherenkov
 telescopes and will make a contribution to study  the  emission mechanism of high energy
 $\gamma$-rays  from  the Crab. 

We report on the first observation of multi-TeV $\gamma$-rays from the Crab with the 
Tibet air shower array.

\section{EXPERIMENT}

 The Tibet experiment, being successfully operated  at Yangbajing in Tibet (4300 m above 
sea level),  now consists of two overlapping arrays (Tibet-II and HD) as described 
elsewhere (\cite{yuda96}). The Tibet-II array consists of  185 scintillation detectors 
of 0.5 m$^2$ each  placed on a 15 m square  grid with an enclosed area of 
36,900 m$^2$, and the HD (high density) array is operated  inside the Tibet-II array to 
detect cosmic ray showers lower than 10 TeV (some of detectors are commonly used in both 
arrays).  This HD array consists of 109 scintillation detectors, each viewed by a  fast 
timing phototube,  placed on a 7.5 m square grid, covering an area of 5,175 m$^2$ (\cite{ame97}). 
A lead plate of 5 mm thickness is placed on the top of each detector in both arrays to 
improve the fast timing data by converting $\gamma$-rays in the showers into electron pairs. 
Using lead converters typically increases the shower size by a factor of about 2 and improves
the angular resolution by about 30 \%. This also tends to flatten the conical shape of the 
shower front fairly (\cite{ame90}).

 The recording 
system is triggered by any fourfold coincidence of the fast timing detectors, which results 
in a trigger rate of about 110 Hz for the HD array. The HD array has been successfully 
operating since 1996 November.   The observation presented here was made with this HD array.

The data used in this search were collected between 1996 November and 1999 May.
The event selection was done by  imposing the following three conditions to the
recorded  data; 1) Each of any four detectors should record a signal of more than 1.25 particles,
2) among the four detectors recording the highest particles, two or more should be inside  
the innermost 9$\times$9 detectors (60~m $\times$ 60~m),
 3) the zenith angle of the incident direction should be less than 45$^\circ$.
After data processing and quality cuts, the total number of events selected was
$1.72 \times 10^9$, and the effective running time was 502.1 days. 

\section{ARRAY PERFORMANCE}

  The vast majority of events detected by the array are initiated by primary cosmic rays
rather than $\gamma$-rays, and therefore it is crucial to reject as much background
as possible.  The background cosmic rays are isotropic and $\gamma$-rays from a source 
are apparently centered on the source direction, so that the reduction of hadronic showers 
must be accomplished via the  good angular resolution of the array.

  In order to achieve a good angular resolution, a study of core-finding techniques and
shower-front curvature corrections has been made (\cite{ame90}). The angular resolution 
of the Tibet array can be well checked  by observing the shadow that the Moon
casts in the cosmic ray flux (\cite{ame93}). Also, the  high flux  of cosmic ray events 
enables us to monitor the system operation by  monthly observation of the Moon's  shadow. 
 Using the HD array, the Moon  shadow was observed with a significance of
15$\sigma$ at the maximum deficit position for all events. From this result, the angular
resolution of the HD array is estimated to be better than 0.9$^\circ$ for all events. 
The resolution increases with increasing
 $\sum \rho$ as $0.8^\circ \times ( (\ge\sum \rho)/20)^{-0.3} \hspace{3mm} (15 <\sum\rho <300)$,
where $\sum \rho$ stands for the sum of the number of shower particles per m$^2$ detected in each 
detector.   

The primary  cosmic rays are mostly  composed of protons and nuclei with positive charge, 
so that those coming from the direction of the Moon are bent by the geomagnetic 
field and the deflection angle, $\Delta\theta$,  of a cosmic ray impinging at normal angle 
on the Earth is calculated as  $\Delta \theta \cong 1.6 (deg) \times (Z/E$), where $E$ 
and $Z$ are the energy of a primary particle in unit of TeV and the atomic number, respectively.
Therefore, the Moon shadow should be found in the direction somewhat away from the 
Moon to the west.  Actually the center of the Moon shadow observed in the events with 
$\sum\rho$ = 15-50 was found at a  position shifted to the west by about 0.3$^\circ$.
On the other hand, we estimated the mean energy of primary protons generating such events at the 
Yangbajing altitude to be about 4.7~TeV according to  the simulations
 (The contribution of helium nuclei
was estimated to be smaller than 20 \%.). 
This value is compatible with  that expected from the observed shift of the shadow and 
the geomagnetic field strength. A more elaborate study of the Moon and  Sun shadows,
using a Monte Carlo technique,  also gives  consistent results with the experiment 
(\cite{suga99}).   Therefore, the observation of the Moon shadow can provide a direct check 
of the relation between shower size and primary energy  when the primary energies are lower 
than 10 TeV.

 Very high energy cosmic rays ( $>20$~TeV), which have a negligible effect from
the geomagnetic field,  should cast a  shadow in the same direction as the Moon. 
Also, it is known that the displacement of the shadow in the north-south direction is almost 
free from the effect of the geomagnetic field. On the basis of these facts, the systematic pointing 
errors of the array can be estimated by examining small deviations of the Moon shadow from the 
correct  position.  It is then found to be smaller than $0.1^\circ$.

 Under the same observation conditions as the Crab,  the trigger efficiencies  of the
array for 3, 4 and 5~TeV $\gamma$-rays and protons are estimated to be 19~\% and 10~\%, 35~\%
and 18~\%, and 47~\% and 26~\%, respectively. That is, the trigger efficiency of $\gamma$-rays is 
about 2 times as large as that of protons when compared at the same energy.
 Hence, observations made by the HD array will partially overlap those of Cherenkov telescopes.

\section{RESULTS AND DISCUSSIONS}

 We used a circular window to search for signals from the Crab.
The ratio of signal from a point source to the isotropic background of hadron primaries
is increased by the good angular resolution of the array. The radius of the window  is chosen to 
optimize  the significance of signals defined as  $N_s/N_B^{1/2}$, where $N_S$ is the 
number of signals and $N_B$ the number of background events, and to contain about 50\% of 
the signals from a source. As shown in the previous section, furthermore, the angular resolution 
of the array depends on the value of $\sum \rho$ in an event. In this analysis, the radii of search 
windows adopted for the events with $\sum\rho >$ 15 (all events), 40, 70  and 130 were 
 0.8$^\circ$, 0.7$^\circ$, 0.5$^\circ$ and 0.5$^\circ$ respectively. 
The signals were searched for by counting the number of events coming from the 
ON-source window.
The background was obtained by averaging over events falling in the ten OFF-source
 windows adjacent to the source. The source window traverses a path in local 
coordinates ($\theta$,$\phi$) through the day, where $\theta$ is the zenith 
angle and $\phi$ the azimuthal angle. In order to avoid the strong zenith angle dependence 
of the background, the OFF-source windows were taken in the $\phi$-directions with the 
same zenith angle $\theta$, except two windows adjacent to the ON-source window. We used the events
with $\theta < 30^\circ$ in the following analysis.

We present the cumulative excess counts of the events with $\sum\rho >$ 15  coming from
the Crab direction as a function of observation time in terms of sidereal day
 in Fig. 1. It is observed  that the excess counts increase linearly with increasing 
observation time, confirming a steady output of multi-TeV $\gamma$-rays from the Crab.

\begin{figure}
\plotone{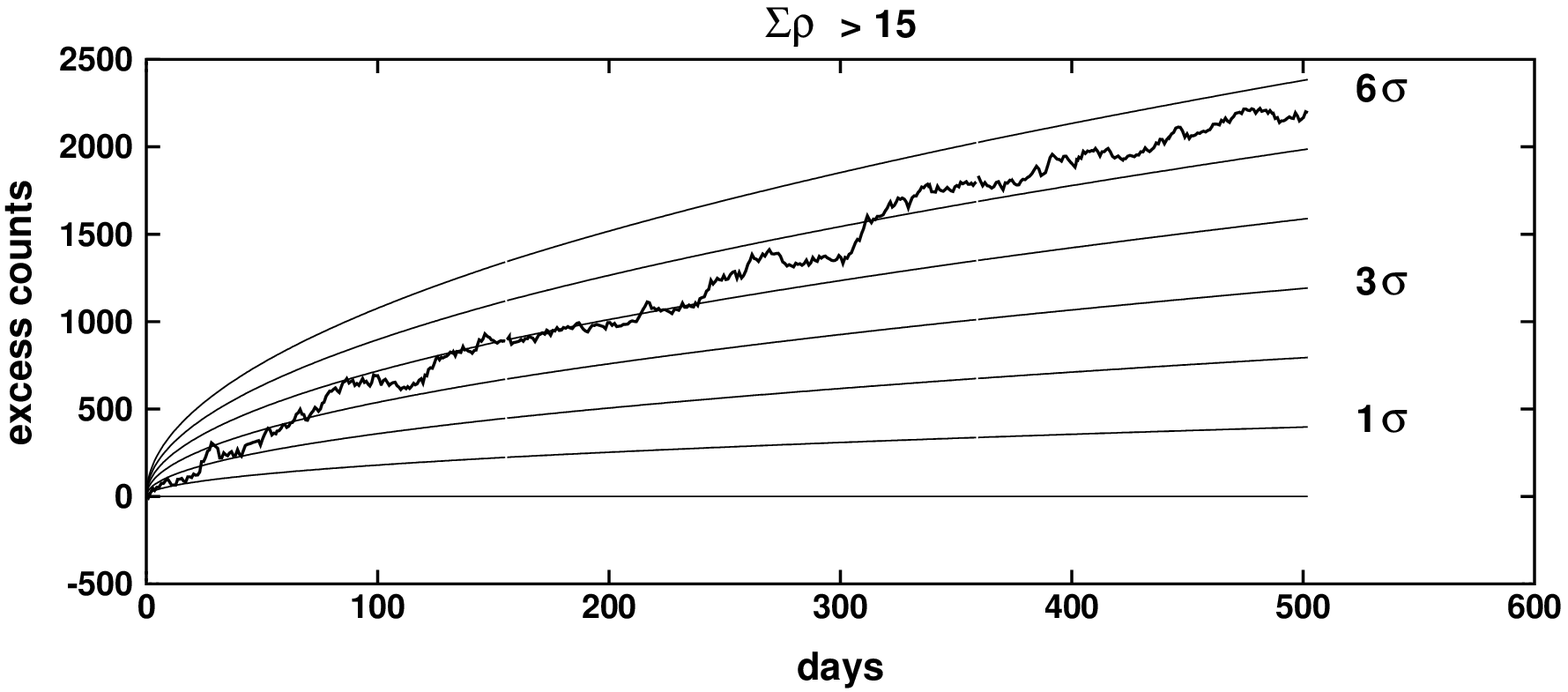}
\epsscale{.7}
\caption{Cumulative excess counts of the events with $\sum\rho >$ 15 
 coming from the Crab direction as a function of  observation time in terms of 
sidereal day. The radius of the search window is taken to be 0.8$^\circ$.  
The solid lines denote the excess counts at respective  significance levels  
starting from 0~$\sigma$, with a step of 1~$\sigma$,  respectively. \label{fig1} }
\end{figure}

\placefigure{fig1}

 Figure 2 shows the opening angle distributions for the events 
with  $\sum\rho >$ 15 and 40, respectively. Here, the opening angle  is defined
as a space angle between the source direction and the event direction. 
The  peak appearing around the origin in the source direction can be attributed to  
$\gamma$-rays from the Crab Nebula. 
 The statistical significances of the excess events with $\sum\rho >$  15, 40,  70 
and 130 are 5.5~$\sigma$,  3.9~$\sigma$, 3.3~$\sigma$  and 3.1~$\sigma$, respectively.

\begin{figure}
\epsscale{.8}
\plotone{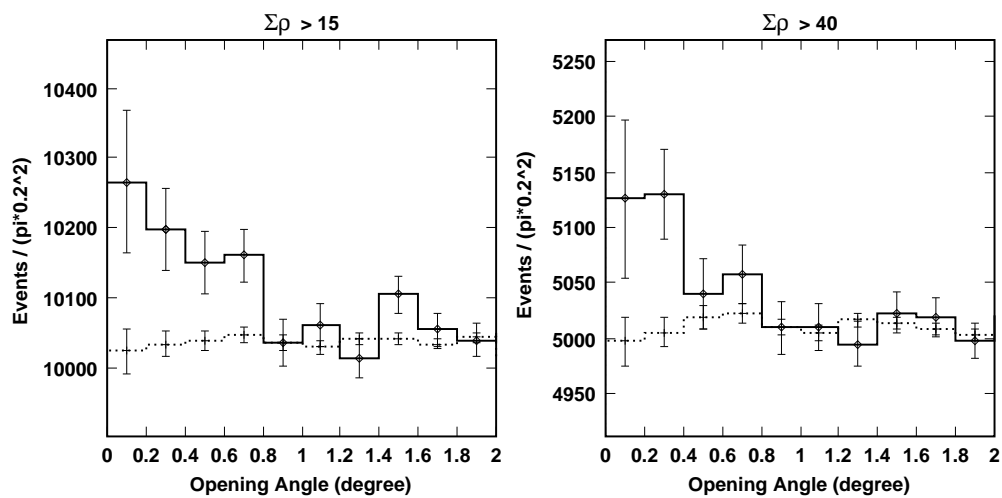}
\caption{Distributions of the opening angles relative to the Crab
 direction for the events with  $\sum\rho >$  15 and  40, respectively.  
Solid and dashed lines  denote ON- and OFF-source data, respectively. \label{fig2}}
\end{figure}

\placefigure{fig2}

In order to estimate  the  Crab spectrum, we calculated the trigger efficiency and the 
collecting area as a function of energy and threshold energy of the array for primary 
$\gamma$-rays  according to the simulation. As discussed in Section 3,
the observations of the Moon shadow   provide an important check of the results obtained 
by assigning energies to all the events. However, the considerable scatter in the energy
of events with the same size will introduce bias into the spectrum because the same size 
event is  enhanced by upward fluctuations of lower energy events and the steeply falling 
spectrum. This  results  in an overestimation of the flux values. This effect must be adequately 
 allowed for in the calculation of the effective collecting area. The  reliability of the 
Monte Carlo for studying the array performance is checked by examining whether or not the
trigger rate of the events can be accounted for by the incoming primary cosmic rays.
 Adopting the primary mass  composition directly measured  by the experiments in the energy 
region lower than 10$^{14}$ eV (\cite{shiba96}) and assuming the heavy enriched composition 
above this region (\cite{ame96}), we calculated the trigger rate of background events by a 
Monte Carlo simulation, taking into account the detector response and  observation conditions.

\begin{figure}
\epsscale{.8}
\plotone{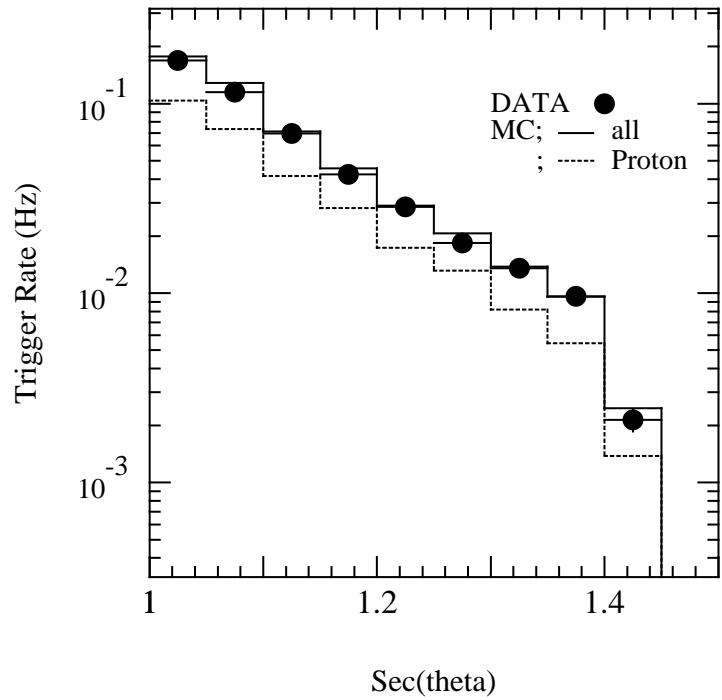}
\caption{Zenith angle dependence of the trigger rates under any four fold
 coincidence in the detectors. The abscissa  is expressed by $\sec \theta$. 
Each of any four detectors records  a signal more than 1.25 particles.  
Simulation results are compared  with the data. Solid and dashed lines show all 
primary particles  and primary protons, respectively. \label{fig3}}
\end{figure}

\placefigure{fig3}

  Shown in Fig. 3 is the zenith angle distribution of the trigger rate of the events 
with $\sum \rho > 15$. We collected the events coming within  a band of 2$^\circ$ width along 
the zenith angle, $\theta$, ranging from 10$^\circ$ (meridian zenith angle of the Crab at 
Yangbajing) to 45$^\circ$.  The experimental data agree well   with the calculated results, 
proving the reliability of our Monte Carlo.  
The simulation also teaches us that the proportion of proton-induced events among all events 
is about 60 \%,  increasing  with increasing zenith angle when  $\sum \rho$ is larger than 50.

  We estimated the $\gamma$-ray spectrum as follows. The primary $\gamma$-rays 
with energies between $E_{\rm min}$  and $E_{\rm max}$  were thrown from the 
direction of the Crab, assuming a differential power-law spectrum of the form 
$E^{-\beta}$ starting at $E_{\rm min}$ = 0.5 TeV,
 which is well below the lowest triggering energy.  The values of  $\beta$ 
and $E_{\rm max}$ were  examined   between 2.4 and 3.0 and between 40 TeV and 100 TeV, respectively.
 Primary $\gamma$-rays, developing  into electron-photon cascade showers (air showers) in the 
atmosphere,  were detected with the array under the same condition as the experiment. 
 Simulated events were divided into respective size bins according to $\sum \rho$
and compared with those of  the experiment.
The energy of $\gamma$-rays is defined as the energy of the maximum flux of simulated events 
in each bin. These steps were repeated until the simulation can well reproduce the
experimental data. We also examined  that the absolute fluxes do not change 
with  the  cut-off energy very much except in the highest energy bin.
 The energy spectrum obtained from this procedure is shown in Fig. 4. We fitted the Crab data 
in the energy range between 3 TeV and 15 TeV by a simple power law, which yields

\begin{equation}
 dJ(E)/dE = (4.61 \pm 0.90) \times 10^{-12} (E/3~TeV)^{-2.62 \pm 0.17} \hspace{5mm}
 {\rm  cm^{-2} s^{-1} {TeV^{-1}}}
\end{equation}

\noindent
where only the statistical errors are quoted. The last three points on  the high energy 
side were  excluded from this fit  because of low significances ($\sim2~\sigma$). 
 Also, the exponent of the spectrum can not be determined  accurately because of the 
small energy range fitted here. The  systematics errors on the flux arise mainly from 
the event selection  which depends upon  the array performance and from the calculations
of the collecting area and the air shower size distribution  by the simulation.
They are estimated to be 13 \% and 8 \%, respectively.  The range of the spectrum
including the statistical and systematic errors is bounded by the dashed lines 
 in Fig. 4.

\begin{figure}
\epsscale{.8}
\plotone{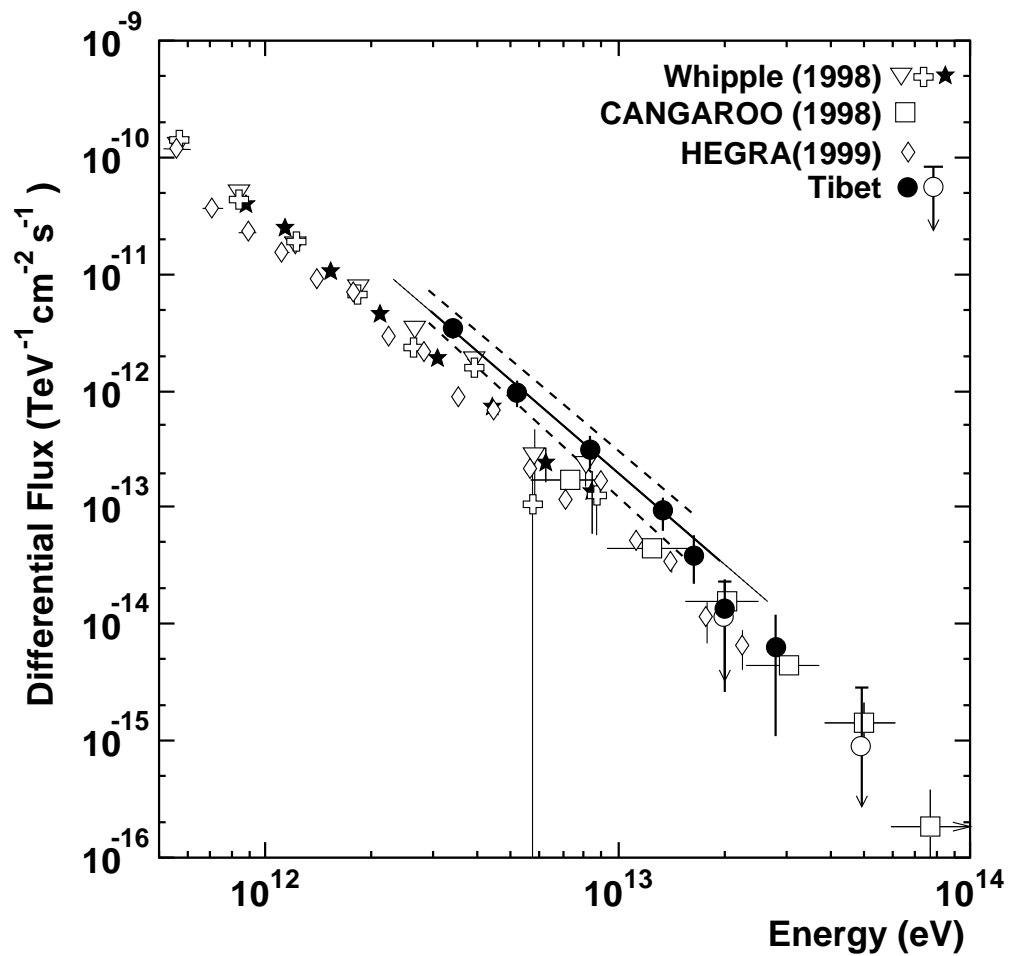}
\caption{Differential spectra of the present result in comparison with those
 of results from the CANGAROO (Tanimori et al. 1998), Whipple (Hillas et al. 1998) 
and HEGRA (Konopelko et al. 1999). The error bars indicate 1~$\sigma$ ranges, 
excluding systematic errors. The full line is the power law fit given in the text.
 For the band shown by the dashed lines, see text.
Upper limit fluxes at the 90~\% confidence level and flux values (open circles),
obtained with the Tibet-II array, are also shown at 20 TeV and 50 TeV.
\label{fig4}}
\end{figure}

\placefigure{fig4}

We also searched for $\gamma$-ray signals from the Crab in the energy region above
10 TeV with the Tibet-II array. Although improving the angular resolution of this array
is still under way, the excess counts are detected at the statistical significances
of $1 \sim 2~\sigma$ level, while marginal yet, at energies around 30 TeV. 
Upper limit fluxes at the 90~\% confidence level are shown in Fig. 4, together with 
other experiments (\cite{hila98} ; \cite{kon99} ; \cite{tani98}).
Recent measurements with imaging Cherenkov telescopes  give  flux values 
 lower  than ours at multi-TeV energies. The CANGAROO data (\cite{tani98}) may not be 
incompatible with ours, but  our data, if anything, may be in favor of a gradual steepening 
of the slope at high energies.
TeV fluxes have sensitivity to probe the spatial variation of magnetic field strength in the nebula, suggesting where the parent electrons exist  in the nebula
(\cite{dej96}). Viewed from this perspective, our data may make a caveat  about  the
estimation of  magnetic field strength (\cite{hila98}). 
This will be  confirmed
by further observations of the spectrum up to high energies with the Tibet array 
in the very near future.

\section{SUMMARY}

 With the high density (HD) air shower array, operating at Yangbajing since 1996, we have
observed  multi-TeV $\gamma$-rays from the Crab Nebula and obtained  the energy spectrum 
of $\gamma$-rays in the energy region above 3 TeV. This is the first clear observation 
of $\gamma$-ray signals  from point sources using a conventional air shower array. 
The success of the air shower array in detecting a signal at the 5-6~$\sigma$ level 
was achieved  by the improvement of the array performance which can be  directly checked 
  by observing the Moon shadow. We have shown that the HD array has  sensitivity to detect 
cosmic ray showers at energies as low as $\sim$3~TeV, 
as well as to observe the Moon shadow with good statistics.  A monthly observation of 
the Moon shadow has made it possible to monitor the angular resolution,  pointing accuracy 
and also the stable   operation of the array over a long period.  

   Our observations  complement telescopes using the atmospheric  Cherenkov technique in the
energy region between 3 TeV and several 10 TeV, so that the present result  may help to reduce 
 the systematic uncertainties in the flux estimates  using imaging atmospheric Cherenkov 
telescopes, and eventually the mechanism of TeV $\gamma$-ray emission from the Crab will be 
further refined. 

The area of the present HD array will be extended by a factor of about five in 1999,
 while its effective area will be increased by a factor of about seven by the reduction
of edge effects.  Then, the Tibet experiment would cover the wide energy range from 3~TeV 
to $\sim$100~TeV, and result in  significantly better statistics and angular resolution 
at high energies.

\acknowledgments
 
 This work is supported in part by Grants-in-Aid for Scientific
Research and also for International Science Research from the Ministry
of Education, Science, Sports and Culture in Japan and the Committee
of the Natural Science Foundation and the Academy of Sciences in
China.

\clearpage

\end{document}